# Liquid crystal films on curved surfaces: An entropic sampling study

D. Jayasri[*†], T. Sairam[‡], K. P. N. Murthy and V. S. S. Sastry
School of Physics, University of Hyderabad, Hyderabad 500 046 Andhra Pradesh, India
[†] Faculty of Physics and Mathematics, University of Ljubljana, 1000 Ljubljana, Slovenia
[‡] Center for Simulational Physics, University of Georgia, Athens, GA 30602, USA

[*]corresponding author: email address: d.jayasri@fmf.uni-lj.si

**Abstract**

The confining effect of a spherical substrate inducing anchoring (normal to the surface) of rod-like liquid crystal molecules contained in a thin film spread over it has been investigated with regard to possible changes in the nature of the isotropic-to-nematic phase transition as the sample is cooled. The focus of these Monte Carlo simulations is to study the competing effects of the homeotropic anchoring due to the surface inducing orientational ordering in the radial direction and the inherent uniaxial order promoted by the intermolecular interactions. By adopting entropic sampling procedure, we could investigate this transition with a high temperature precision, and we studied the effect of the surface anchoring strength on the phase diagram for a specifically chosen geometry. We find that there is a threshold anchoring strength of the surface below which uniaxial nematic phase results, and above which the isotropic fluid cools to a radially ordered nematic phase, besides of course expected changes in the phase transition temperature with the anchoring strength. In the vicinity of the threshold anchoring strength we observe a bistable region between these two structures, clearly brought out by the characteristics of the corresponding microstates constituting the entropic ensemble.

## I. Introduction

Confined liquid crystals have the potential to yield macroscopic states with interesting, and sometimes unexpected, equilibrium director structures; and transitions among them have been attracting attention both from the point of view of understanding the role of resulting free energy landscapes in stabilizing the different phases, as well as, in favourable cases, of exploring new applications based on such new director structures. It is now appreciated that by a subtle variation of experimental conditions (like temperature, aligning direction and its intensity, length scale of confinement, external stimulus like field and radiation, etc.) novel transitions can be triggered initially at the surface, often leading to bulk alignment itself [1]. One of the early studies on thin nematic films in contact with substrate surface inducing homeotropic boundary conditions was made by Sheng [2] based on Landau-de Gennes theory. His work, and several studies that followed, (see e.g. [3]) showed that the nematic-isotropic transition temperature ($T_{NI}$) is sensitive to the thickness of the nematic film confined between two substrates. Subsequently, several authors have employed molecular dynamics as well as Monte Carlo simulation techniques to probe the transitional behaviour of a liquid crystal confined between boundaries that induce conflicting orders via a variety of inter-molecular interactions [4]. Recent DNMR studies [5] show that there exists a narrow co-existence region of bulk-like and surface-induced ordering when observed as a function of surface thickness.

The present work is aimed at simulating possible deirector structures that emerge in a thin liquid crystal film (comprising of uniaxial molecules) deposited on a spherical substrate. Recent work [6] on spherical shells of nematic medium (embedded suitably between two emulsions) indicated the rich defect structures that can be induced in the medium under specific boundary conditions. In this context, we consider a thin liquid crystal film formed on a small enough sphere, which permits a competition between the surface-induced radial ordering and inherent preference of the medium to align uniaxially. The outer layer of the droplet experiences a free boundary condition. The relative influences of these two antagonistic mechanisms can be tuned by introducing a variable anchoring strength at the surface ($\varepsilon_s$) on one hand, and on the other by tuning the curvature of the surface and the number of molecules participating in the film via the substrate radius ($r$) and film thickness ($d$). We focus on the regime wherein this competition leads to a bistablity between the two distinct structures. In Section-II we introduce the Hamiltonian model of the system, and simulational details. Section-III presents the results of the computations, discussing the temperature variation of various observable properties for differing surface-induced effects.

## II. Hamiltonian model and details of simulations:

We model the interaction among the participating uniaxial liquid crystal molecules in terms of the Lebwohl-Lasher potential [7] which is widely used to capture the essential thermodynamic features of the isotropic-to-nematic phase transition. This lattice model assumes that each lattice site is occupied by a headless vector (appropriate to the apolar symmetry of the liquid crystal system) and the interaction is restricted to only the nearest neighbours. The model Hamiltonian accounting for all such interactions between neighbouring sites hosting unit vectors (representing the molecular orientations in principle, or more appropriately local directors formed due to a small cluster of molecules) is given by

$$U = -\sum_{i,j} \epsilon_{ij} \left( \frac{3}{2} \left( u_i \cdot u_j \right)^2 - \frac{1}{2} \right) \quad (1)$$

We consider the interaction strength among the different sites to be the same (no bond disorder), thereby setting $\epsilon_{ij} = \epsilon$, and use this as the energy unit for specifying the temperature in reduced units ( $T^* = k_B T/\epsilon$ ). The above summation is restricted to nearest neighbours. This Hamiltonian in bulk sample is known to yield an isotropic-to-nematic phase transition temperature $T_{NI} = 1.1232$ [8].

We construct the desired shell of the liquid crystal film by considering a cubic lattice of adequate dimensions to contain both the spherical substrate and the film of required thickness. In order to mimic the substrate $\mathbb{S}$, we consider the centre of the cube to be coincident with that of the sphere, and treat all the unit vectors residing on the lattice points inside the (jagged) sphere of radius $r$ to be oriented in directions parallel to the respective radial vectors, and fixed. The film of the liquid crystal $\mathbb{F}$ spread over the substrate is identified with the lattice sites which lie within the two (jagged) spheres with radii $r$ and $(r + d)$. The orientations of the unit vectors within $\mathbb{F}$ vary during the Monte Carlo simulation, leading to microstates spanning the configuration space of the film. The lattice sites outside the radius $(r + d)$ are not part of the system under consideration and they have no interactions with the liquid crystal sites inside (corresponding to free boundary conditions on the outer layer of the liquid crystal film). While the interaction strength $\epsilon$ is the same for all the terms in eqn. (1) involving liquid crystal sites, the corresponding value $\epsilon_s$ between the liquid crystal and substrate sites is made variable, allowing for the flexibility to study the effect of the surface. The objective of the simulation is to build equilibrium ensemble of microstates of this system at different temperatures and compute useful physical properties (like the orientational order in the film, average energy, specific heat (at constant volume), and nematic susceptibility as averages over these ensembles.

We initially experimented with the values of $r$ and $d$ so as to subject the system to comparable competition between the two opposing ordering mechanisms, and found that a choice of $r = d = 3$ is optimal to be able to observe some interesting results. It may be noted that the importance of the respective mechanisms depend on the curvature at the spherical substrate and the number of liquid crystal sites participating in the interaction of eqn. (1). These preliminary studies also showed that the transition temperature shifts towards higher temperature as $\epsilon_s$ is increased, and there is a threshold value of this parameter on either side of which the system goes over to either a radially ordered spherical shell (higher $\epsilon_s$) or to a uniaxial nematic film. In the neighbourhood of this threshold value, the simulations pointed out to the possibility of coexistence of the two differently ordered structures. Keeping this in view, we employed a more recent Monte Carlo sampling method which forces the system to perform a random walk (in its configuration space) which is sampled uniformly with respect to its energy. This has the advantage of making the sampling insensitive to the energy barriers that might be present in the system particularly near the phenomena like the above. Thus the data presented in this work is derived from such (energy-uniform or non-Boltzmann) collection of microstates by reconstructing the canonical ensembles using established reweighting methods [9, 10].

Entropic sampling and equilibrium properties:

Recently, with the advent of efficient algorithms to build multi-canonical ensembles in particular the Wang-Landau (WL) algorithm [10], it has become possible to compute representative density of states (DoS) of a given model system through Monte Carlo sampling method. WL algorithm was modified to make it more readily applicable to systems with continuous degrees of freedom like liquid crystals with continuous reorienational degrees of freedom [11]. Based on this procedure, we first obtained the DoS of the film under consideration, and subsequently made the system perform an energy-uniform random walk by biasing the walk against its own DoS. This leads to a collection of microstates, which form the entropic ensemble of the system. The canonical ensembles are then extracted simply by simultaneously applying two biasing probabilities to the microstates, one according to the DoS and the other due to the assumed statistical distribution (Maxwell-Boltzmann) corresponding to the temperature under consideration [so-called reweighting]. The assignment of the resultant probability to the microstates in the entropic ensemble completes the construction of the canonical ensemble, in so far as their contribution to the desired physical property is concerned. One of the useful outcomes of this sampling is that any desired temperature resolution can be achieved by appropriate reweighting from the entropic ensemble (once collected), and secondly all the microstates in the (much larger) entropic sample in principle contribute to the canonical ensemble (to varying degree of importance of course) leading to a relatively noise-free simulation result. We adopt this procedure [11] in reporting the following equilibrium properties of the film under different boundary conditions.

Considering the geometry and the boundary conditions applied to the film, two order parameters are of interest here: the uniaxial order $S_A$ describing the degree of orientational ordering of the molecules along a resultant director $\boldsymbol{n}$, and the radial order $S_R$ quantifying the compliance on the part of the liquid crystal molecules to the imposed boundary condition at the substrate surface. $S_A$ is obtained using the standard procedure of computing the average ordering matrix of the unit vectors at the liquid crystal lattice sites, and finding its eigen values and eigen vectors. The eigen values are directly related to the amount of uniaxial order $S_A$ and the phase biaxiality (if any) [12], while the eigen vectors provide the average ordering directions. In particular, the maximum eigen value is connected with the uniaxial order $S_A$, and its direction defines the director $\boldsymbol{n}$. $S_A$ of a microstate then is a measure of the deviation of the phase from the isotropic symmetry, and is given by

$$S_A = \left(\frac{1}{N}\right) \sum_{i=1}^{N} \frac{1}{2} \left(3\cos^2(u_i \cdot n) - 1\right) \tag{2}$$

Here, N is the total number of liquid crystal sites in the film. The radial order is computed, as per the following equation, by first calculating projections of each of the unit vectors at the liquid crystal sites along the local radial direction and then taking an average (of a suitable function) over them.

$$S_R = \left(\frac{1}{N}\right)\sum_{i=1}^{N}\frac{1}{2}\left(3\cos^2(u_i \cdot R) - 1\right) \qquad (3)$$

Here, $R$ is the unit vector along the radial direction at the site $i$ bearing the unit vector $u_i$. Both these orders (computed for each microstate) are further averaged over the equilibrium ensemble.

The fluctuations in the microstate energy over the ensemble are related to the specific heat (at constant volume) $C_v$, providing useful signatures of the phase transition.

## III. Results and discussion:

In all these simulations the temperature is varied between 0.5 and 1.5 (reduced units) and equilibrium ensembles are constructed in intervals of 0.01. The anchoring strength $\epsilon_S$ is varied between 0 and 2, initially in steps of 0.1 in order to determine its threshold value (as discussed earlier), and thereafter in steps of 0.01 bracketing this threshold value. Figs. 1 and 2 show the variation of $S_A$ and $S_R$ with $T$ for different values of $\epsilon_S$. The two variations complement each other in depicting the temperature variation of the director structure, modulated by the variable surface interaction. The value of $T_{NI}$ increases with increasing $\epsilon_S$, as seen by the peak positions of the $C_v$ profile at the corresponding anchoring values (Fig.3). The broadening of these peaks indicates softening of the weak first order isotropic-to-nematic transition, expected from Lebwohl-Lasher Hamiltonian (in bulk samples). While the positive shift with increased (antagonistic) surface perturbation is to be expected, the variation of the order parameters is quite interesting. Fig. 1 shows that the axial order is reasonably high at lower values of $\epsilon_S$ and in fact is comparable to bulk values at the corresponding temperatures. It is also seen that the formation of the uniaxial structure, in terms of its onset with temperature, is however delayed as $\epsilon_S$ is increased from 0 to say 1.5. In this $\epsilon_S$ range, the radial order (which did not exist at $\epsilon_S$ = 0) grows initially with decrease of temperature (below the corresponding $T_{NI}$, as per the $C_v$ peak position), progressively encompassing more temperature region as $\epsilon_S$ is increased gradually. In the corresponding experimental situation, the profiles of $S_A$ (Fig. 1) show that the onset of this order is that much delayed. The scenario thus corresponds to a paranematic phase (isotropic phase with surface-induced non-zero radial order) undergoing a phase transition to a 'nematic' phase but with a predominantly radial director distribution. The gradual build up (made effective below the temperature-driven transition) of the axial order eventually forces the system to transit to an essentially uniaxial character. This is borne out by the profiles in this $\epsilon_S$ range from Figs. 1 and 2. Thus the temperature at which the transition takes place ($T_{NI}$) is not the same as the temperature near which the uniaxial phase actually forms – it requires further cooling for the elastic property of the medium to come into play to overcome the (radial) disordering effect. Within the resolution adopted in the present work, this continues till the value of $\epsilon_S$ is about 1.55, the so-called threshold value. Above this value the surface anchoring seems to be too strong for the elastic response of the nematic medium to force a uniaxial order, and the medium develops a monotonically increasing $S_R$ with decrease in temperature. The variation of the average energy with temperature at the corresponding $\epsilon_S$ values is shown in Fig. 4, indicating the gradual progression of the transition towards higher temperature and its gradual softening. This sample in a way thus provides an interesting example of a paranematic-to-nematic transition in the presence of an inhomogeneous 'field', while its perturbation can be controlled and the system response can be quantified, still retaining the inhomogeneous character.

Keeping the curious behaviour of the two competing orders in view, we examined the nature of the entropic ensembles collected during this study (several million microstates distributed fairly uniformly with respect to the sample energy). We computed the radial and axial orders of all the microstates and plotted them as a function of sample energy at different anchoring strengths. The resulting graphs depict the distribution of microstates (representatively) on the $S_A - E$ and $S_R - E$ diagrams. Fig. 5 shows such distributions very near the anchoring threshold value (for $\epsilon_S$ = 1.55, 1.56, and 1.57). The plots corresponding $\epsilon_S$= 1.55 have many interesting features: The structures immediately below the transition have discernible radial order which grows steadily with decrease of energy (corresponding to decrease of temperature) till a specific value, below which the radial order suddenly drops yielding to axial order. At very low energies, the uniaxial order is significantly high (about 0.85), while the radial order is about 0.2 indicating the remnant radial order in the surface layer of the liquid crystal. This scenario changes at $\epsilon_S$ = 1.56: the number of microstates corresponding to the high axial order diminishes considerably at low enough energies, while the distribution of microstates extends to higher radial order, and relatively more populated. Curiously for both the cases, there are complementary low probable branches: i.e. there is a preponderance of highly ordered states with radial symmetry, but with a sparsely distributed low order branch as well. This is correlated in the opposite way with the uniaxial ordered states, as expected. This coexistence of differently ordered macrostates (evidenced by the distribution of microstates) shows that the system is exhibiting bistability with respect to these two distinct structures, the control parameter (for a given low enough energy and hence temperature) being $\epsilon_s$. The corresponding distributions plotted for $\epsilon_s$ = 1.57 show further progression of the well ordered radial structures of the director into low energy regions. Finally, within the errors that creep in when only a finite number of microstates can be sampled in any simulation, one can see from Fig. 5 the possibility that perhaps the system may as well be having microstate-rich distributions in certain ranges of either of the order parameters, interspersed by regions with much lower probability of occupancy.  Such a scenario in this toy model of a tunable frustrated system is perhaps to be expected, alluding to the rugged free energy profile of a glassy system.

**Acknowledgements**

The simulations were carried out at the Centre for Modelling Simulation and Design (CMSD) at the University of Hyderabad. DJ would like to thank the Department of Science and Technology, New Delhi, for the Research Fellowship through the HPCF project at CMSD.

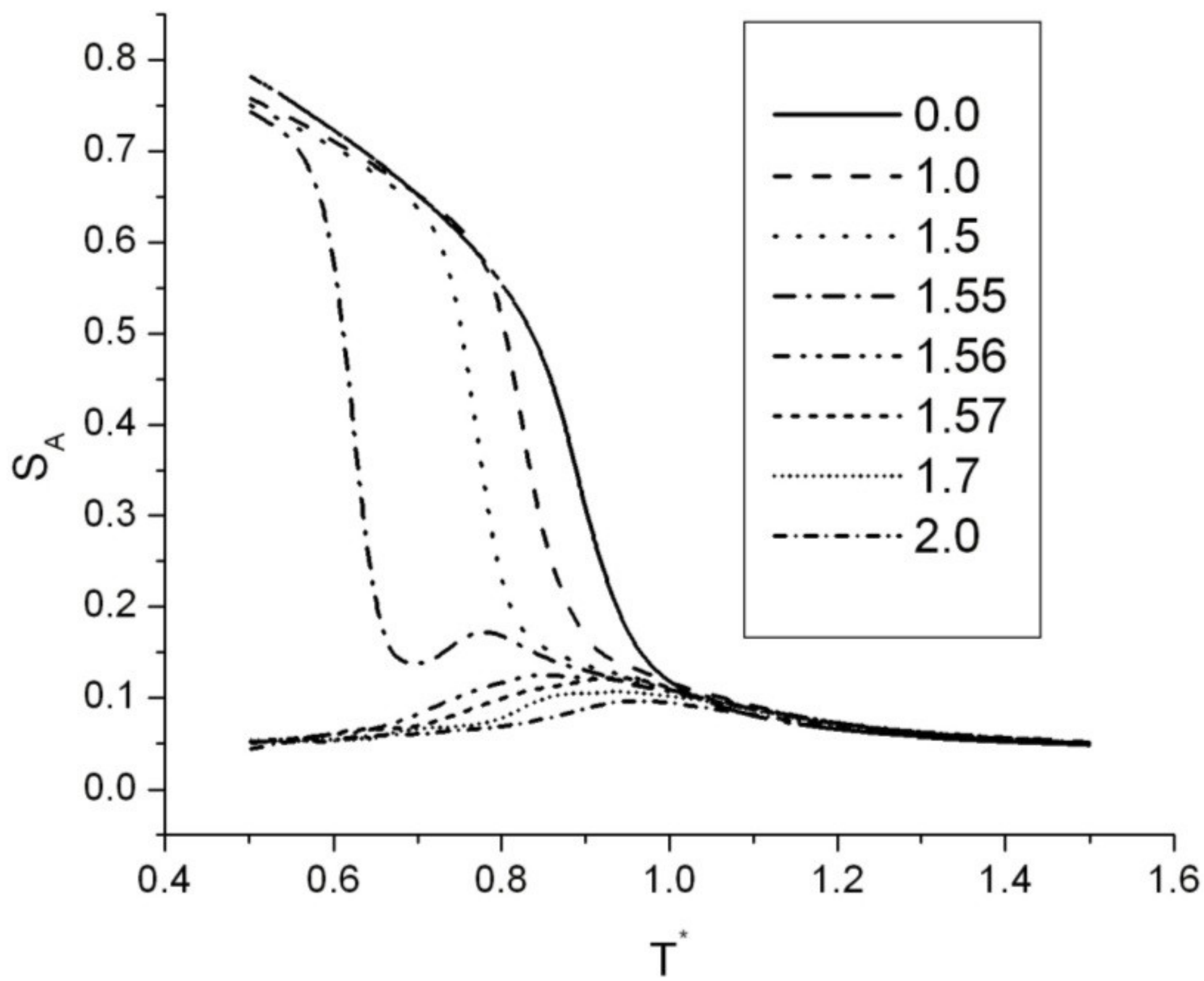

Fig 1: Axial orientational order parameter $S_A$ for various values of anchoring strengths $\varepsilon_s$= 0.0 to 2.0.

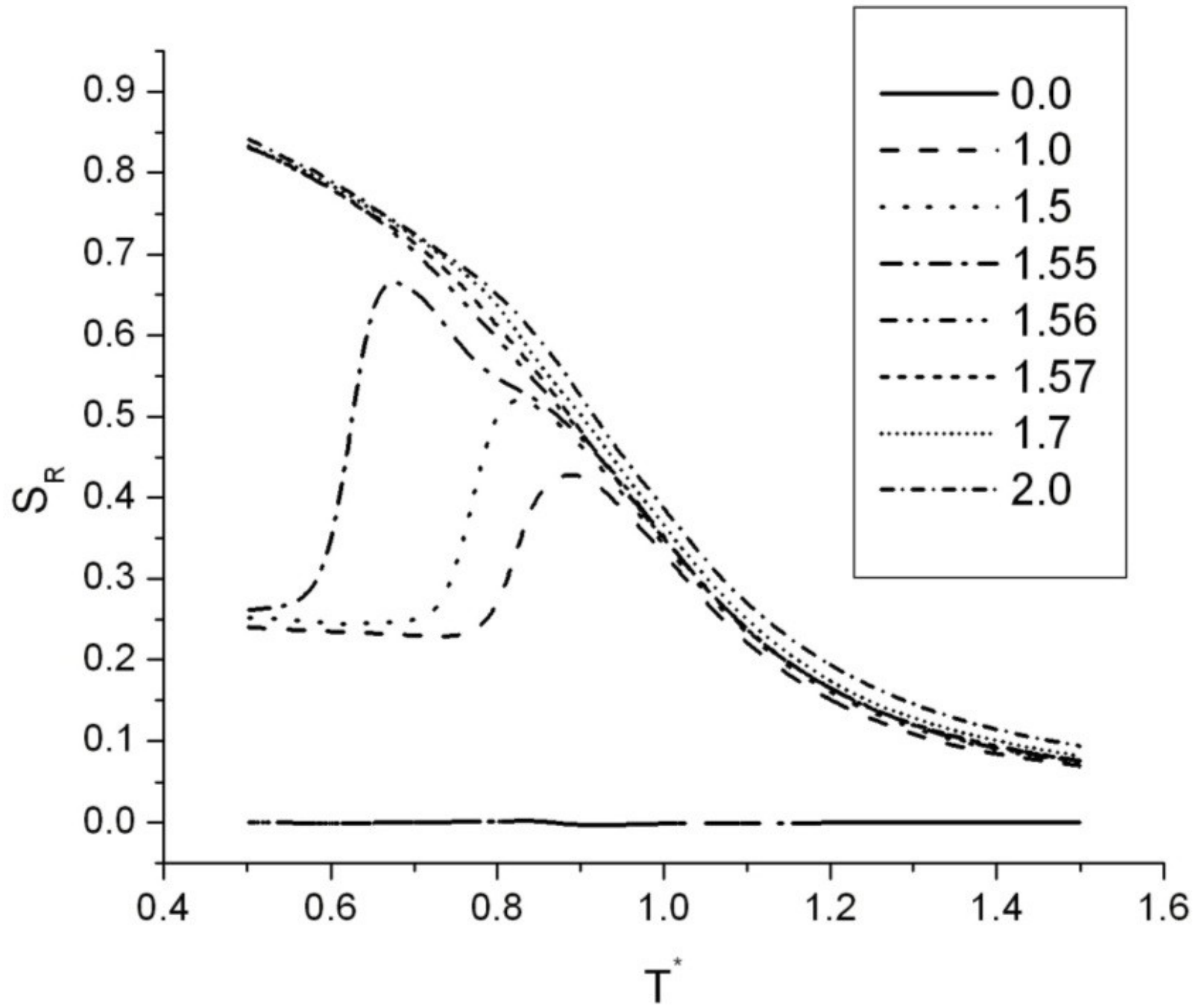

Fig 2: Radial orientational order parameter $S_R$ for various values of anchoring strengths $\varepsilon_s$= 0.0 to 2.0.

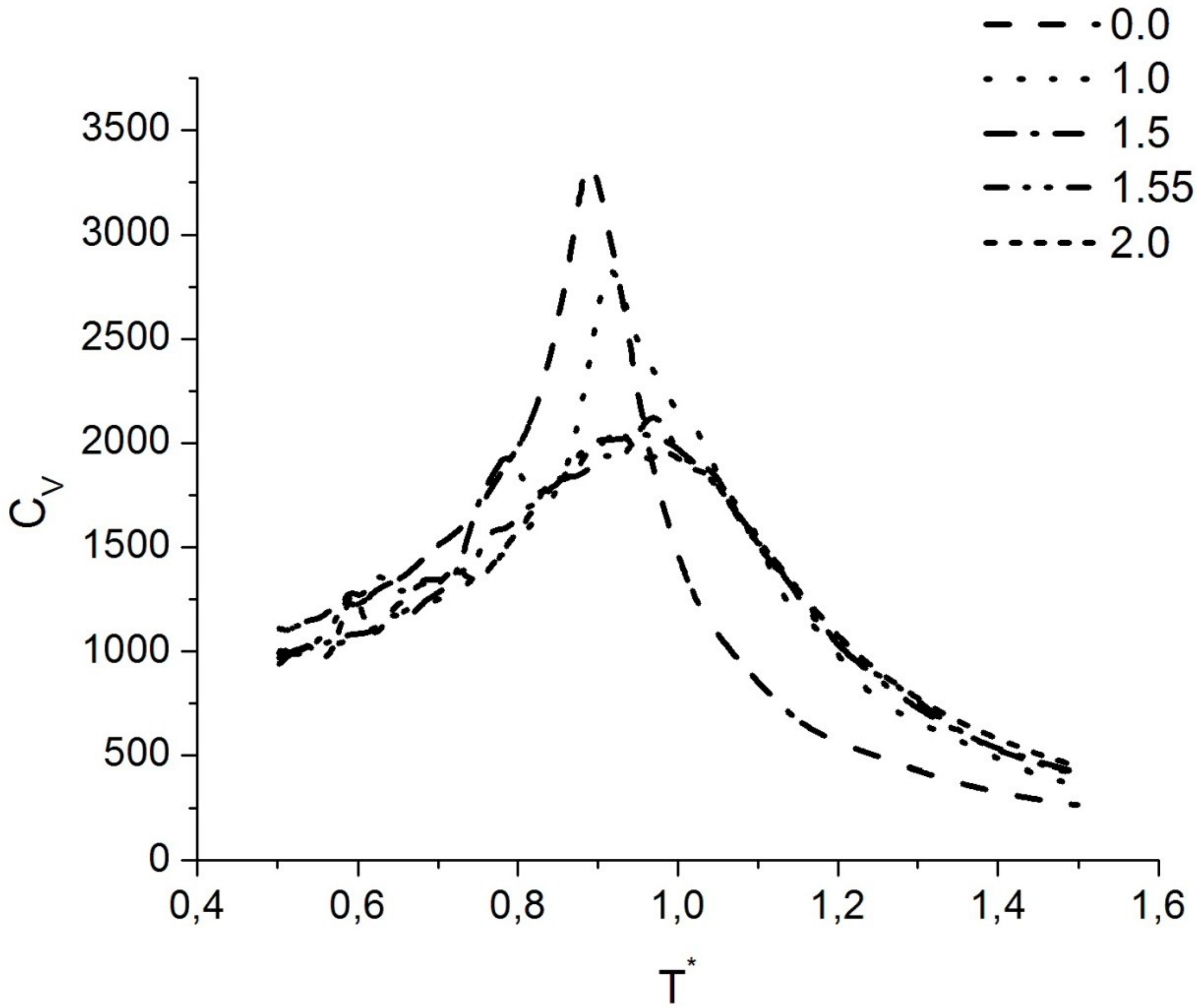

Fig 3: Specific heat versus temperature for anchoring strengths $\varepsilon_s$= 0.0 to 2.

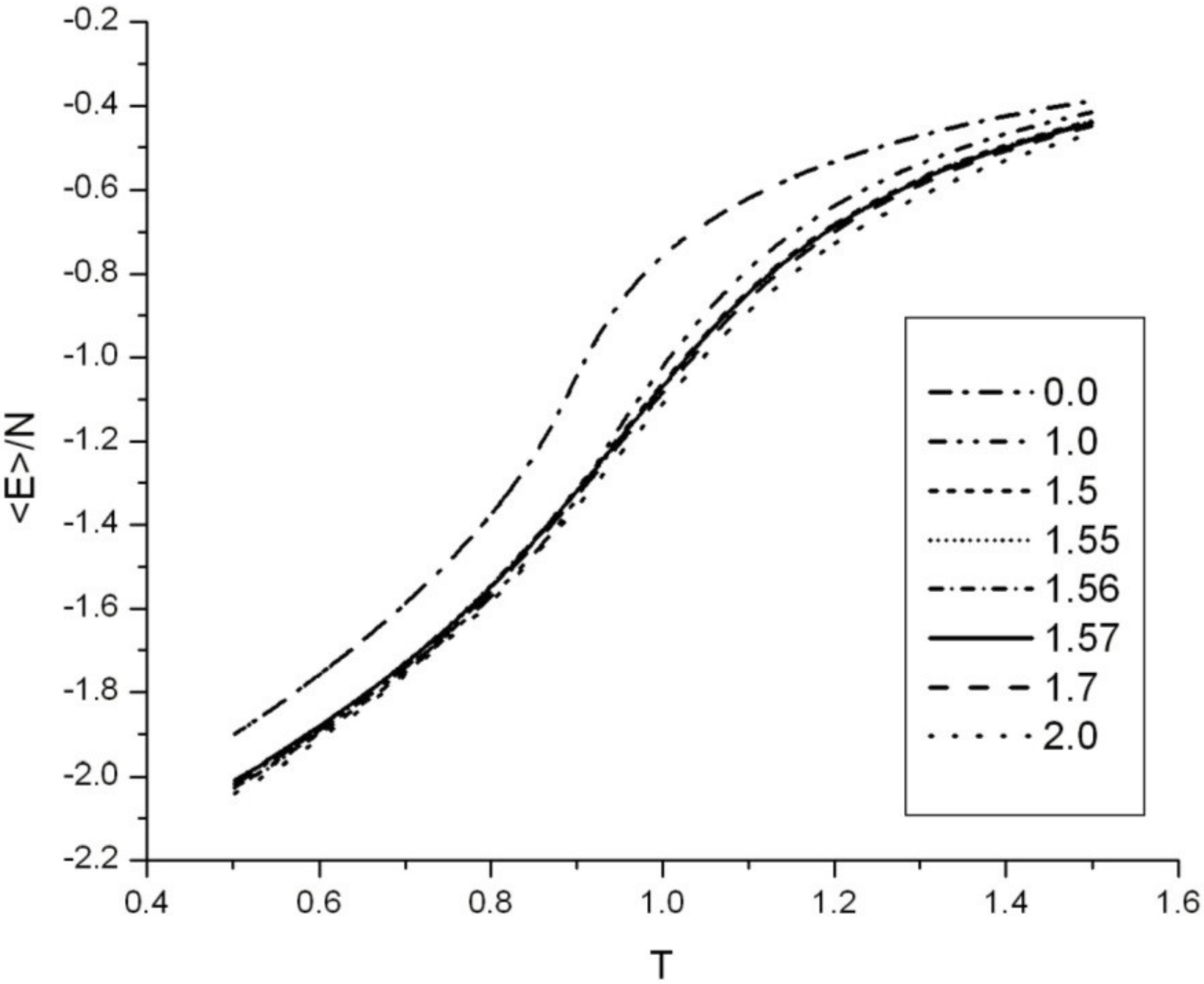

Fig 4: Average energy versus reduced temperature for anchoring strengths $\varepsilon_s$= 0.0 to 2.0.

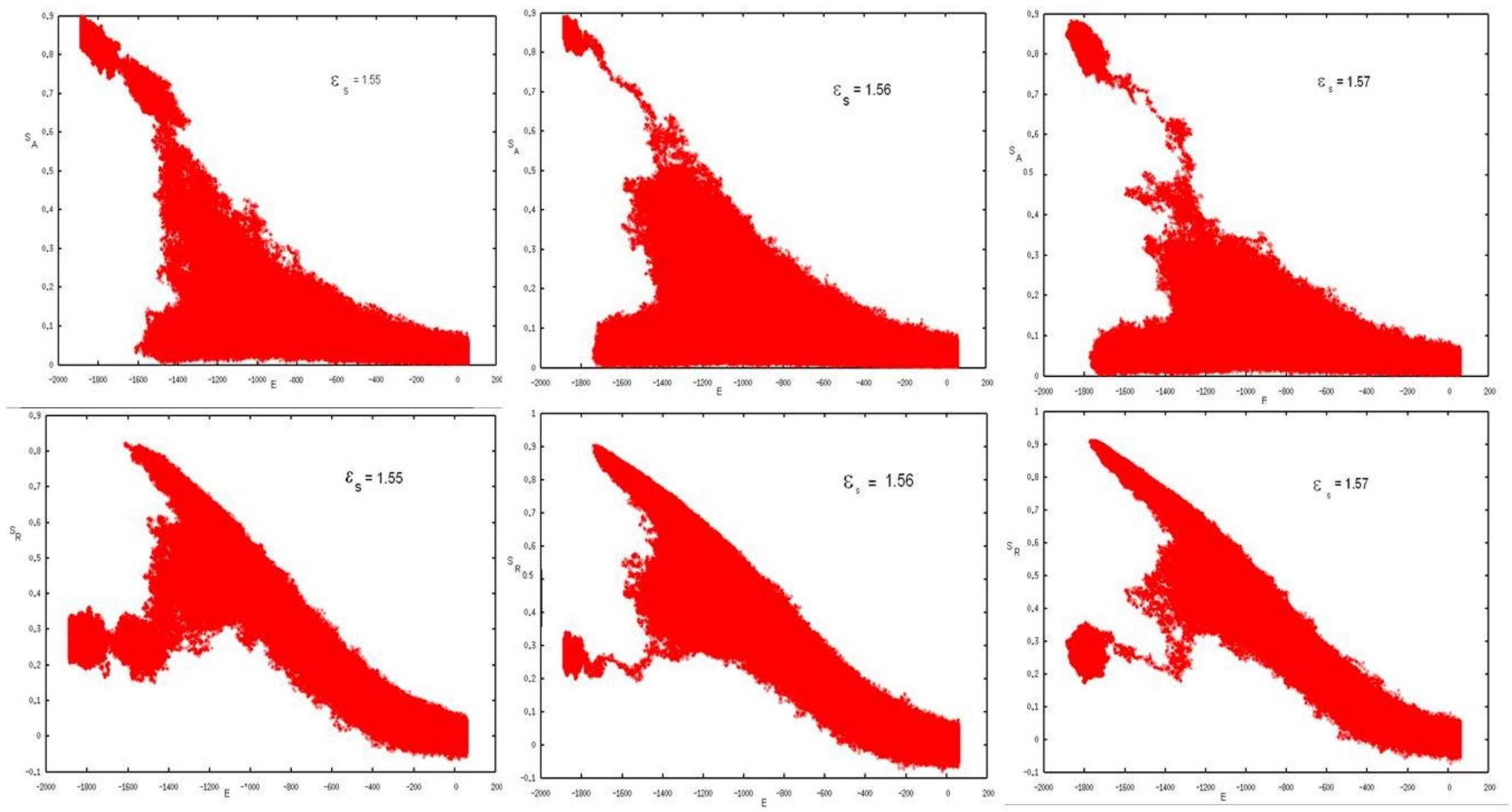

Fig 5: Non-Boltzmann ensembles collected during the production run for $\varepsilon_s$= 1.55, 1.56 and 1.57.